\documentclass{iaus}
\usepackage{graphicx}

\title[The mid-term periodicities in sunspot areas]
{The mid-term periodicities in sunspot areas}

\author[R. Getko]
{Ryszarda Getko$^1$}  

\affiliation{$^1$Astronomical Institute, University of Wroclaw, Wroclaw, Poland 
\\ email: {\tt getko@astro.uni.wroc.pl} }

\pubyear{2008}
\volume{257}  
\pagerange{1--3}
\jname{Universal Heliophysical Processes}
\editors{N. Gopalswamy \& D. Webb, eds.}
\begin{document}

\maketitle

\begin{abstract}

The sunspot area fluctuations for the northern and the southern hemispheres of the Sun over the epoch of 12 cycles (12-23) are investigated. 
Because of the asymmetry of their probability distributions, the positive and the negative fluctuations are considered separately. 
The auto-correlation analysis of them shows three quasi-periodicities at 10, 17 and 23 solar rotations. 
The wavelets gives the 10-rotation quasi-periodicity. 
For the original and the negative fluctuations the correlation coefficient between the wavelet and the auto-correlation results 
is about 0.9 for $90\%$ of the auto-correlation peaks. For the positive fluctuations it is also 0.9 for $70\%$ 
of the peaks. For $90\%$ of cycles in both hemispheres the auto-correlation analysis 
of negative fluctuations shows that two longer periods can be represented as the multiple 
of the shortest period. For positive fluctuations such dependences are found for more than $50\%$ of cases.
 

\end{abstract}

\section{Introduction}

  In the last decades, the intermediate quasi-periodicities of many solar activity tracers have been discussed.
The about 12-rotation periodicity identified \cite[Krivova \& Solanki (2002)]{kri} for sunspot data during 1749-2001. 
It was prominent during times of stronger activity, whereas it diminished and sometimes faded into the background during weak cycles. 
\cite[Getko (2006)]{get} found it in both high and low activity periods 
for the monthly Wolf numbers during cycles 1-22 and for the group sunspot numbers during cycles 5-22.
Two longer quasi-periodicities at 17 rotations and at 23 rotations were found 
in many solar activity parameters from the bottom of the convection zone to the atmosphere.
More up-to-date review is by \cite[Obridko and Shelting (2007)]{ob}. Here I present results from a statistical study of these periodicities 
in the sunspot areas during cycles 12-23. It enables one to deduce the mean length of the time period between strong fluctuations.

\section{A detailed analysis of quasi-periodicities}

 \begin{figure}
   \centerline{\includegraphics[width=3.4in]{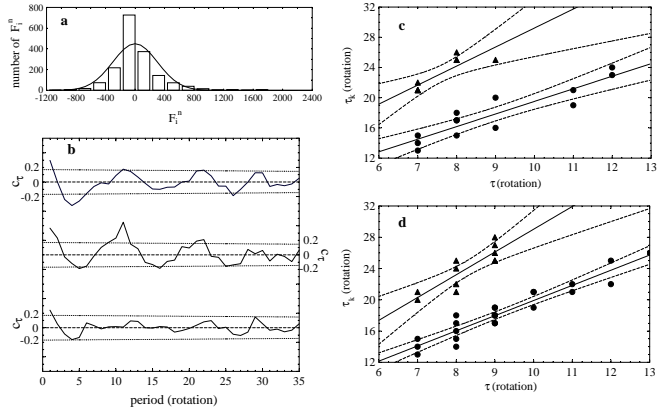}}
    \caption{ \rm \bf{a}  \rm Histogram of $\{F_i^{n}\}$ with a fitted Gaussian. 
                  \bf{b} \it{Top:} \rm Auto-correlation function ($c_\tau$) of $\{F_i^{n}\}$ for cycle 18. 
\it{Middle:} \rm Same as for the upper curve, but for $\{F_i^{n-}\}$.
    \it{Bottom:} \rm Same as for the upper curve, but for $\{F_i^{n+}\}$.
     The dotted lines represent two standard errors of each $c_\tau$ function.  
    \bf{c} \rm Dependence of $\tau_2$ and $\tau_3$ on $\tau$  for positive fluctuations. 
Lower solid curve represents the regression line for the points $(\tau, \tau_2)$ (dots). 
Upper solid curve represents the regression line for the points $(\tau, \tau_3)$ (triangles). 
Dashed lines represent the 95 per cent confidence interval for each regression line.
 \bf{d} \rm Same as for  \bf{c}\rm, but for negative fluctuations.}
   \label{i1}
   \end{figure}
I consider the daily sunspot areas for the northern hemisphere $(D_l^n)$, and the southern hemisphere $(D_l^s)$
for solar cycles 12-23 available at the National Geophysical Data Center (NGDC)
(http://solarscience.msfc.nasa.gov/greenwch/). 
For the $i$-th Carrington rotation I evaluate the mean sunspot area for the northern hemisphere $(S_i^n)$:
$S_i^n=\frac{1}{L} \sum_{l=1}^{L}{D_l^n}$,
where $L$ is the number of days for the $i$-th rotation. I define the fluctuation $(F_i^n)$ of the mean sunspot area 
$(S_i^n)$ from the smoothed mean sunspot area:
$F_i^n=S_i^n-\overline{S_i^n}\quad \mbox{for} \quad i=1,\ldots, N$, where $\overline{S_i^n}=\frac{1}{13} \sum_{j=i-6}^{i+6}{S_i^n}$. 
Each of the time series $\{F_i^{n}\}$ and $\{F_i^{s}\}$ contains $N=1706$ elements. 
Both have almost the same probability distributions. 
Fig. \ref{i1}a shows the histogram of $\{F_i^{n}\}$ with a fitted Gaussian.
The Kolmogorov-Lilliefors and Shapiro-Wilk tests reject the hypothesis of normality for them.
Because each distribution has positive skew, the positive and the negative fluctuations are considered separately.
For the northern hemisphere they can be defined as follows:\\
\begin{minipage}{6.1cm}$F_i^{n+}=\left\{\begin{array}{r@{\quad\mbox{where}\quad}l}
  0 & F_i^n\le 0 \\ F_i^n & F_i^n>0
  \end{array} \right. \mbox{and}$
\end{minipage}\begin{minipage}{5cm}
$F_i^{n-}=\left\{\begin{array}{r@{\quad\mbox{where}\quad}l}
  0 & F_i^n> 0 \\ F_i^n & F_i^n\le 0
  \end{array} \right.$ 
\end{minipage}
\begin{minipage}{4.0cm}
for $i=1,\ldots,N.$
\end{minipage}\\

It is known that for a time series which contains Gaussian white noise 
and a sinusoidal component a probability distribution is symmetric and the auto-correlation functions $(c_\tau )$ of that time series, 
of its positive fluctuations and of its negative fluctuations should be the same.
The functions $c_\tau$ of $\{F_i^n\}$, $\{F_i^{n+}\}$ and $\{F_i^{n-}\}$ for one solar cycle
are different (Fig. \ref{i1}b). The functions $c_\tau$ of $\{F_i^{n}\}$ and $\{F_i^{n-}\}$ for cycle 18 have 
the significant global maxima at $\tau=11$ rotations and smaller maxima at $\tau_1=k*\tau$ for $k=2$ and $3$. 
In $54\%$ of 24 cases (12 cycles in each hemisphere) the  functions $c_\tau$ of the original fluctuations have significant 
maxima for $\tau\in [7, 13]$. In $30\%$ cases the maxima for such $\tau$ belong to 
the interval $[1\sigma, 2\sigma]$. For positive fluctuations this contribution is $50\%$ and $46\%$ respectively.
For negative fluctuations in $92\%$ cases the maxima at $\tau\in [7, 13]$ are significant.
The mean value of all $\tau\in [7, 13]$ for which the maxima are significant is approximately 10 rotations for 
all three  fluctuation groups. I also consider the $c_\tau$ maxima for $\tau\in [14, 19]$ and $\tau\in [20, 27]$.
For more than $50\%$ of cases the positive fluctuations create the auto-correlation peaks for which the periods 
$\tau_1\approx 17$ and $\tau_2\approx 23$ can be represented as $\tau_k\approx k*\tau$ where $\tau\in [7, 13]$ and 
$k=2$ or $3$. For  each of  $k$ the points $(\tau, \tau_k)$, the regression line (solid) and 
the $95\%$ confidence interval for each line (dotted) are shown in Fig. \ref{i1}c. For $k=2$ the correlation coefficient for 13 points 
is 0.91, for $k=3$ it is 0.86 for 8 points. For the negative fluctuations the strong dependence between 
the considered periodicities was found in $\sim 90\%$ for $k=2$ and in $\sim 50\%$ for $k=3$ (Fig. \ref{i1}d). 
For $k=2$ the correlation is 0.95 for 23 points and for $k=3$ it is 0.91 for 11 points. 
It is important to add that the $c_\tau$ values at $\tau >27$ are not reliable because of the solar cycle length.
\begin{figure}
\centerline{\includegraphics[width=3.4in]{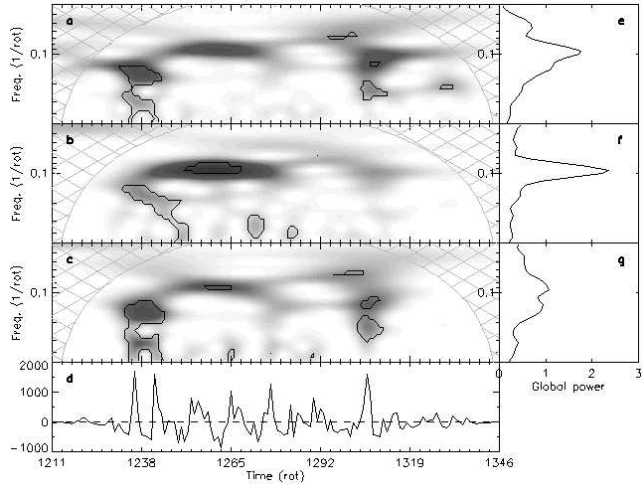}}
\caption{\bf {a-c:} \rm  Wavelet power spectra of \bf{a:} \rm $\{F_i^{n}\}$, \bf{b:} \rm$\{F_i^{n-}\}$ and \bf{c:} \rm $\{F_i^{n+}\}$ 
mapping a time-frequency evolution of about 10-rotation periodicity. 
Top values of wavelet power are denoted by gradual darkening. 
Black contours denote significance levels of 95 per cent for detected peaks. A cone of influence is marked by the dashed region. 
\bf{e-g:} \rm  Corresponding global wavelet power spectra.  \bf{d:} \rm Time series $\{F_i^{n}\}$ for cycle 18.
} 
\label{i2} 
\end{figure}

I also applied the Morlet wavelet (\cite[Torrence \& Compo 1998]{tor}) to three fluctuation time series for each of 24 cases. 
Fig. \ref{i2} shows  the normalized wavelet maps for cycle 18. Black contours denote the 95 per cent significance level for detected peaks.
The wavelet map of $\{F_i^{n}\}$ (top) shows two significant peaks (at $\tau\approx 6$ and $8$). 
They are mainly created by three strong fluctuations at the begining and at the end of the high activity period. 
During the remaining part of this period a rised power is at $\tau = 11$ rotations. 
Moreover, the integrated spectrum (right) shows the maximum at $\tau = 11$ and confirms the auto-correlation results. 
For $\{F_i^{n-}\}$ (middle) the peak at $\tau=11$ is well detected with 95 per cent level, extends in time during the high activity period 
and dominates the integrated spectrum. The map of $\{F_i^{n+}\}$ (bottom) is similar to the map of $\{F_i^{n}\}$, 
but the peak at $\tau = 11$ is significant. The global spectrum shows two almost the same peaks at $\tau = 11$ and $8$. 
Such an analysis was done for all 24 cases. The auto-correlation and the wavelet results are similar for 
$\tau\in [7, 13]$ (the correlation between them is 0.9 for $87\%$, $92\%$ and $72\%$ of the auto-corre\-la\-tion peaks of $\{F_i^{n}\}$, 
 $\{F_i^{n-}\}$ and $\{F_i^{n+}\}$ respectively).

These results could indicate that the 10-rotation quasi-period is dominant. \cite[Getko (2004)]{get1} showed that 
large activity complexes were responsible for strong sunspot number fluctuations. Thus, the time between strong fluctuations of 
toroidal magnetic flux in the tachocline could be on the order of 7-13 rotations. 
It is also well known that two solar hemispheres show certain hemispheric asymmetries in their solar-cycle features. 
However, two-sample Kolmogorov-Smirnov test shows that the 10-rotation quasi-periods evaluated for each of 12 solar cycles 
in each hemispheres do not differ. Two longer quasi-periods at about $17$ and $23$ rotations could be treated as subharmonics 
of the 10-rotation quasi-period (Figs.  \ref{i1}c and  \ref{i1}d). This facts could explain a wide range of periodicities 
in various solar indices at all levels from the tachocline to the Earth.
  
\section{Conclusions}
  
\begin{enumerate}
\item For both hemispheres the probability distributions of fluctuations are similar and have an asymmetry which means that 
there are more negative than positive fluctuations.
\item The auto-correlation analysis of the original, the positive and the negative fluctuations prefers three quasi-periods: around 10, 17 and 23 rotations. 
The wavelet maps show one dominant quasi-period at about 10 rotations.
\item For $90\%$ of solar cycles in both hemispheres the auto-correlation analysis of negative fluctuations gives peaks 
for which the period $\tau_2\approx 17$ rotations can be represented as $\tau_k\approx k*\tau$ where $\tau\in [7, 13]$ and 
$k=2$. For $k=3$ a such dependence are reliable in $50\%$ of considered cases.
For positive fluctuations such dependences are found for more than $50\%$ of solar cycles in each hemispheres. 
\end{enumerate}

\end{document}